\documentclass[largeformat]{interact}
\usepackage[T1]{fontenc}
\usepackage[utf8]{inputenc}
\usepackage{lmodern}
\usepackage{etoolbox}
\makeatletter
\patchcmd{\@maketitle}
  {\receivedfont{\bfseries ARTICLE HISTORY\\}\@received\par}
  {}{}{
    \PackageError{manuscript}{Could not remove article history}{Check interact.cls.}
  }
\makeatother
\usepackage{natbib}
\bibpunct[, ]{(}{)}{;}{a}{}{,}

\usepackage[hidelinks]{hyperref}
\hypersetup{pdftitle={Looking Forward to Constructing Better Causal Comparisons},pdfauthor={Augusto Cerqua}}

\begin{document}

\title{Looking Forward to Constructing Better Causal Comparisons}
\author{%
  \name{Augusto Cerqua\thanks{CONTACT Augusto Cerqua. Email:
    \href{mailto:augusto.cerqua@uniroma1.it}{augusto.cerqua@uniroma1.it}}}
  \affil{Department of Social Sciences and Economics, Sapienza University of Rome, Rome, Italy}
}
\maketitle

\noindent{\small This version: 24/10/2026\par}
\vspace{8pt}

\begin{abstract}
Constructing credible comparison groups is a central challenge in observational causal inference. Conventional approaches assess comparability through observed characteristics or pre-treatment outcome trajectories, although causal identification depends on unobserved untreated potential outcomes after treatment. We propose a forward-looking framework that selects or weights controls using prognostic scores: predictions, based on pre-treatment information, of the outcomes units would experience without treatment. These scores target either untreated outcome levels or changes, depending on the comparability condition underlying the causal design. They can be estimated using post-treatment outcomes observed for untreated units or exclusively from historical data. To construct these scores, we leverage tabular foundation models, whose extensive pretraining enables them to capture complex relationships among predictors and effectively extract their predictive information, even when application-specific samples are limited. Crucially, predictions guide comparison-group construction, whereas observed outcomes determine treatment effect estimates. We derive an exact decomposition of estimation error into prognostic-score imbalance and differences in forecast errors, clarifying why neither predictive accuracy nor score balance alone guarantees causal identification. The National Supported Work application reveals that stronger pre-treatment covariate balance does not necessarily yield estimates closer to experimental benchmarks, highlighting the practical relevance of prognostic-score comparisons.
\end{abstract}


\section{A forward-looking comparison}

In causal inference, the purpose of a comparison group is to approximate what would have happened to treated units in the absence of an intervention. Researchers usually assess suitable controls through pre-treatment characteristics and outcome histories. Matching seeks similarity in observed covariates, while difference-in-differences (DiD) commonly uses pre-treatment trajectories to assess the plausibility of parallel trends. In both cases, the causal question concerns outcomes that remain unobserved after treatment begins.

These diagnostics are informative, but they cannot establish the assumptions required for causal identification. To identify the average treatment effect on the treated (ATT), matching relies on the conditional independence assumption (CIA), requiring untreated potential outcomes to be independent of treatment assignment conditional on observed covariates, together with sufficient overlap. DiD instead relies on the parallel trends assumption (PTA), requiring treated and control units to experience the same average outcome changes in the absence of treatment. Both assumptions are fundamentally untestable in observational studies: covariate balance cannot establish conditional independence, just as similar pre-treatment trends cannot guarantee parallel trends in the post-treatment period. Consequently, even perfect balance or identical pre-treatment trends would not be sufficient to establish the validity of the corresponding identifying assumptions. As emphasized by \citet{KahnLang2020}, parallel pre-treatment trends are neither necessary nor sufficient for the validity of the parallel trends assumption in the post-treatment period. More recently, \citet{Ghanem2026} show that the informativeness of pre-trend comparisons depends on the underlying treatment selection mechanism.\footnote{Throughout the analysis, we assume no anticipation and no interference between units. Although interference represents an additional threat to causal identification \citep{Cox1958}, we set this issue aside to focus on the assumptions underlying the comparability of treated and control units.}

Building on \citet{Hansen2008}, this paper proposes an alternative approach to constructing comparison groups based on prognostic scores, namely forecasted outcomes in the absence of treatment. These forecasts can be obtained either by exploiting historical relationships between pre-treatment outcomes and observed covariates or by training predictive models on the observed post-treatment outcomes of untreated units.  Prognostic scores provide a forward-looking criterion for selecting or weighting control units based on their expected outcomes in the absence of treatment, rather than relying exclusively on similarities in observed characteristics or historical outcomes. The underlying intuition is that units with substantially different observed characteristics or historical outcome trajectories may nevertheless experience similar changes in the outcome of interest in the absence of treatment, whereas units that are similar in terms of relevant pre-treatment covariates may exhibit different outcome trajectories under the untreated regime. Importantly, the relevant dimension of comparability depends on the identification strategy. For comparisons in outcome levels, controls are selected to resemble treated units in their forecasted untreated outcome levels, whereas for DiD, comparability is assessed in terms of forecasted untreated outcome changes. In both cases, forecasting makes the target of comparability explicitly forward-looking, focusing on the post-treatment period in which the causal effect is to be estimated.\footnote{Although this paper focuses on matching and DiD, we believe that the forward-looking approach proposed here could also be valuable for other causal inference methods. In regression discontinuity designs, prognostic scores could potentially help extend treatment effect estimation beyond the immediate vicinity of the cutoff by identifying comparable units farther from the threshold. In instrumental-variable approaches, outcome-based prognostic adjustments could potentially improve estimation efficiency.}

Importantly, the forward-looking approach does not replace the identifying assumptions underlying matching or DiD. Its main contribution is to use predicted untreated outcomes to identify suitable comparison units for each causal design, while relying on their observed outcomes to estimate treatment effects. We further show how estimation errors can be decomposed into differences in predicted outcomes and differences in forecast errors between treated and control units. To evaluate the approach in an applied setting, we revisit the National Supported Work (NSW) demonstration, which provides an experimental benchmark against which observational estimates can be evaluated. This application allows us to assess whether selecting controls based on predicted untreated outcomes yields estimates closer to the experimental benchmark than conventional propensity-score methods. Using tabular foundation models (TFMs), we construct prognostic scores and apply nearest-neighbor matching and entropy balancing to select or weight comparison units. We find that combining DiD with entropy balancing on prognostic scores trained using observed post-treatment outcomes of untreated controls yields an ATT estimate particularly close to the experimental benchmark.

\section{Prognostic scores and predictive models}

The concept of prognostic scores was formally introduced by \citet{Hansen2008}, building on earlier contributions such as the multivariate confounder score proposed by \citet{Miettinen1976}. While propensity scores summarize the relationship between observed covariates and treatment assignment \citep{RosenbaumRubin1983}, prognostic scores capture their relationship with potential outcomes under a reference treatment condition. Hansen shows that, under suitable assumptions, prognostic scores can reduce the dimensionality of observed covariates while preserving the information needed for causal adjustment. This provides a theoretical foundation for constructing comparison groups based on their expected outcomes rather than requiring similarity across all observed characteristics. Formally, a prognostic score is a function of covariates that renders the potential outcome under the reference condition independent of the original covariates, conditional on the score. For average treatment effects, the conditional expectation of the untreated potential outcome provides a natural prediction target.\footnote{In this paper, we use estimated conditional means as working prognostic scores, without requiring them to satisfy the full distributional sufficiency condition underlying Hansen's definition.}

The use of historical data to construct prognostic scores also has practical precedents. For example, \citet{Wyss2015} used information from previous patients to predict health risks and identify comparable groups of patients receiving different treatments. \citet{Smith2015} used pre-treatment event histories to construct time-dependent prognostic scores in an evaluation of liver transplantation. The importance of prognostic scores for observational study design is further emphasized by \citet{AikensBaiocchi2023}, who introduce assignment-control plots to visualize the joint distribution of propensity and prognostic scores, highlighting how these two dimensions capture different aspects of comparability between treated and control units. Building on this literature, our framework uses forecasted untreated outcomes to select or weight comparison units, adapting the prognostic score to the specific requirements of matching and DiD.

Recent advances in tabular foundation models (TFMs) have made flexible and accurate prediction increasingly feasible, even when application-specific training samples are limited. Through extensive pretraining, these models can capture complex relationships and adapt to new datasets with little or no task-specific training. A prominent example is TabPFN, which is pretrained on synthetic prediction tasks and uses the supplied dataset as context to generate predictions.\footnote{Unlike conventional machine learning models, TabPFN is pretrained on millions of synthetic tabular prediction tasks, learning to recognize and exploit complex relationships between predictors and outcomes across different datasets. When applied to a new dataset, it uses observed predictor–outcome pairs as context to generate predictions for new observations, without updating its pretrained parameters. This process, known as in-context learning, enables flexible prediction without conventional task-specific model training and can be particularly advantageous when application-specific data are limited.} \citet{Hollmann2025} document its strong performance on small and medium-sized tabular datasets, while the more recent TabPFN~3.5 technical report extends the evaluation to additional settings, including temporally structured and grouped data \citep{Jager2026}. Recent benchmarks show that state-of-the-art TFMs now consistently outperform traditional machine learning methods, including random forests and XGBoost, across the vast majority of tabular prediction tasks \citep{Hollmann2025,Jager2026}. These developments open new opportunities for forecasting untreated potential outcomes. Although strong predictive performance cannot guarantee the validity of the assumptions required for causal identification, combining TFMs with our forward-looking approach offers a promising way to exploit the available information more effectively and construct comparison groups that better reflect the expected outcomes in the absence of treatment.

\section{Prognostic scores and identification}
\label{sec:framework}

\subsection{Choosing the forecasting target}

Consider observations at two pre-treatment dates, $-1$ and $0$, followed by treatment before date $1$. Let $D_i$ indicate treatment and $Y_{i1}(0)$ denote the potential outcome at date $1$ in the absence of treatment, defined for every unit regardless of its treatment status. This outcome is observed for untreated units but remains counterfactual for treated units.

A conventional implementation of the prognostic-score approach introduced by \citet{Hansen2008} exploits observed post-treatment outcomes among untreated units. Specifically, a predictive model $\widehat g_0(x)$ can be estimated using their observed outcomes at date $1$ and pre-treatment covariates $X_i$. The resulting predictions, $\widehat g_0(X_i)$, provide estimates of untreated potential outcomes for both treated and control units, allowing comparisons based on their expected outcomes in the absence of treatment. The main advantage of this approach is that predictions are informed by outcomes observed during the evaluation period, avoiding the need to extrapolate historical relationships into the future. However, constructing prognostic scores using post-treatment outcomes may introduce overfitting concerns, particularly when the same control observations are subsequently used for causal estimation.

We explore an alternative strategy in which prognostic scores are constructed entirely from pre-treatment information. Specifically, a predictive model learns the relationship between information available at date $-1$ and the outcome observed at date $0$. Applying this relationship to information available at date $0$ then yields a forecast $\widehat m_i$ of $Y_{i1}(0)$ for every unit. When additional pre-treatment periods are available, the model can exploit multiple historical transitions and benefit from temporal validation. The distinctive feature of this alternative strategy is that prognostic scores and comparison weights are determined without using outcomes from the evaluation period. This separates the construction of the comparison group from the subsequent estimation of treatment effects, while preserving the possibility of exploiting observed post-treatment outcomes to account for shared forecasting errors. The trade-off is that the approach requires at least two pre-treatment observations and sufficiently stable predictive relationships over time, particularly when forecasting over longer horizons.

The algebraic decompositions developed below apply equally to prognostic scores constructed exclusively from pre-treatment information and to those incorporating evaluation-period outcomes from untreated units. In both settings, they clarify the role of score balance in causal identification and highlight the importance of forecast-error comparability and sufficient overlap between treated and control units. Moreover, the relevant comparison score depends on the causal design, irrespective of how predictions are obtained. For comparisons in outcome levels, the objective is to identify controls with predicted untreated outcome levels similar to those of treated units. For DiD, comparability instead concerns predicted untreated outcome changes relative to the last pre-treatment observation. These two criteria can be expressed as follows:

\begin{equation}
s_i^{\mathrm{Level}}=\widehat m_i,
\qquad
s_i^{\mathrm{DiD}}=\widehat m_i-Y_{i0}.
\label{eq}
\end{equation}

The superscripts $\mathrm{Level}$ and $\mathrm{DiD}$ distinguish between scores based on predicted untreated outcome levels and changes, respectively, rather than the method used to construct comparison groups. Both scores can be used to select or weight control units through matching, stratification, or reweighting methods such as entropy balancing. The resulting comparison groups can then be used to estimate treatment effects either from post-treatment outcome levels, as commonly done in propensity-score matching applications, or from outcome changes using DiD.

\subsection{Forecast errors and identification}

Forecasts determine how controls are selected or weighted; observed outcomes determine the estimated effect. To see why this distinction matters, let $\mathcal T$ and $\mathcal C$ denote treated and control units, and let control weights $\omega_j$ sum to one. For any unit-level quantity $a_i$, define the difference between the treated mean and the weighted control mean as
\begin{equation}
\Delta_{\omega}a
=\frac{1}{n_T}\sum_{i\in\mathcal T}a_i
-\sum_{j\in\mathcal C}\omega_j a_j,
\qquad n_T=|\mathcal T|.
\label{eq:treatment_effects}
\end{equation}
Write the untreated outcome as $Y_{i1}(0)=\widehat m_i+r_i$, where $r_i$ is the forecast error. Let $\tau_T$ be the sample average treatment effect among treated units. Under no anticipation, the level-based and DiD treatment effect estimates satisfy the exact decompositions
\begin{equation}
\begin{aligned}
\widehat\tau_{\mathrm{Level}}-\tau_T
&=\Delta_{\omega}s^{\mathrm{Level}}+\Delta_{\omega}r,\\
\widehat\tau_{\mathrm{DiD}}-\tau_T
&=\Delta_{\omega}s^{\mathrm{DiD}}+\Delta_{\omega}r,
\end{aligned}
\label{eq:forecast_error_decomposition}
\end{equation}
where $\widehat\tau_{\mathrm{Level}}=\Delta_{\omega}Y_1$ and $\widehat\tau_{\mathrm{DiD}}=\Delta_{\omega}(Y_1-Y_0)$.

Equation~\eqref{eq:forecast_error_decomposition} clarifies the relationship between prognostic-score balance and causal identification by separating estimation error into two components: differences in predicted untreated outcomes and differences in forecast errors between treated and weighted control units. Exact balance on the relevant score eliminates the first component, leaving unbiasedness dependent on whether forecast errors are equal in expectation across the two groups. When score balance is approximate, as under nearest-neighbor matching, both components contribute to estimation error. The validity of the resulting causal estimates therefore depends on the comparability of forecast errors, alongside sufficient overlap to construct suitable comparison groups. Importantly, neither high predictive accuracy nor score balance alone guarantees these conditions.

Using observed control outcomes can help account for forecasting errors shared by treated and control units. For example, an unexpected shock affecting all units equally cancels out because the weights assigned to treated and control units each sum to one. By contrast, shocks that affect the two groups differently do not cancel. Forecast-based comparisons therefore require forecast errors to be comparable, on average, between treated and weighted control units, rather than accurate predictions for every unit. This is a key advantage of combining predictive modeling with observed controls: forecasts guide the selection of suitable comparison units, while their observed outcomes help account for common deviations from predicted untreated outcomes.

This framework is closely related to recent developments in difference-in-differences (DiD) that incorporate outcome modeling into causal estimation, although prediction plays a distinct role in our approach. \citet{SantAnnaZhao2020} combine models of untreated outcome changes with propensity score weighting to construct doubly robust estimators, while \citet{Callaway2021} extend outcome-regression, inverse probability weighting, and doubly robust approaches to settings with multiple periods and staggered treatment adoption. Similarly, \citet{Borusyak2024} develop an imputation-based DiD estimator that predicts untreated potential outcomes using models fitted to observations not exposed to treatment. More directly related to our focus on comparison-group construction, \citet{hong2012} propose a prognostic-score-based DiD strategy that subclasses units based on paired predictions of pre-policy and untreated post-policy outcomes. Our implementation uses a single scalar comparison score targeting either untreated outcome levels or changes. We consider predictions obtained from evaluation-period outcomes of untreated controls and predictions constructed exclusively from pre-treatment information. These scores guide matching or reweighting, while treatment effects are estimated from observed outcomes. The implementation thus differs in its score construction and comparison procedures, rather than in the general principle of combining prognostic adjustment with DiD. More broadly, our framework extends beyond DiD by offering a flexible approach to comparison-group construction across different causal inference designs, leveraging recent advances in tabular foundation models (TFMs) to make better use of the available pre-treatment information.

\subsection{Validation and diagnostic checks}

Evaluating the proposed approach requires assessing both predictive performance and the credibility of the resulting comparison groups. When sufficient pre-treatment data are available, historical validation can assess forecasting accuracy using earlier periods. Examining the overlap in prognostic scores between treated and control units, together with the concentration of control weights, helps determine whether suitable comparisons can be constructed without relying excessively on a small number of controls. Pre-treatment covariate balance also remains informative, as units with similar predicted untreated outcomes may still differ substantially in observed characteristics.

When sufficiently long pre-treatment histories are available, an additional validation strategy consists of in-time placebo tests, as commonly employed in the synthetic control and DiD literature. The idea is to assign a fictitious treatment date before the actual intervention and replicate the entire estimation procedure using only information that would have been available at that point. Predictive models are trained on earlier observations, comparison units are selected or weighted based on the resulting forecasts, and placebo treatment effects are estimated over subsequent periods that still precede the actual intervention. Since no treatment has yet occurred, estimated effects should be close to zero if the procedure successfully reconstructs the untreated outcome trajectories of future treated units. Repeating this exercise across different placebo dates can provide further insights into the stability and reliability of the proposed comparisons.

These placebo exercises go beyond evaluating predictive accuracy by assessing whether forecasts and observed controls can jointly reproduce the outcomes of treated units before treatment. However, successful placebo tests do not guarantee that the same comparability will hold after treatment begins. Moreover, as emphasized by \citet{CerquaLettaPinto2026}, validation with panel data must account for temporal and cross-sectional dependencies to avoid data leakage and overly optimistic assessments of predictive performance. Ultimately, while these diagnostics can strengthen confidence in the proposed approach, they cannot verify its identifying assumptions. This is a fundamental limitation of observational causal inference, as the relevant counterfactual outcomes are inherently unobservable.

\section{Empirical application}
\label{sec:nsw_application}
\enlargethispage{4pt}

\subsection{The benchmark and the comparison population}

The National Supported Work (NSW) demonstration provides an unusually informative setting for assessing observational comparison designs. Conducted in the United States during the 1970s, the program offered subsidized employment and training to individuals facing substantial barriers to labor market participation. Its randomized design provides an experimental benchmark against which estimates obtained using nonexperimental comparison groups can be evaluated. In his seminal study, \citet{lalonde1986} exploited this setting by replacing randomized controls with individuals drawn from large household surveys, including the Current Population Survey (CPS). He found that conventional nonexperimental estimators frequently failed to reproduce the experimental earnings effects, often yielding estimates that were highly sensitive to model specification and sometimes substantially different in magnitude or even sign. These findings raised fundamental concerns about the reliability of observational program evaluations and established the NSW data as a widely used benchmark for assessing alternative methods of constructing credible counterfactuals. A more encouraging perspective emerged from \citet{dehejia1999} (DW), who revisited LaLonde's analysis and showed that carefully specified propensity-score matching methods, combined with attention to pre-treatment characteristics and common support, could produce estimates substantially closer to the experimental benchmark. Their findings highlighted the importance of carefully constructing comparison groups to obtain reliable causal effect estimates from observational data. Subsequently, \citet{smith2005} showed that the performance of propensity-score matching in the NSW data is sensitive to the analysis sample and score specification, with difference-in-differences matching performing best among the estimators they examined. Building on this debate, we revisit the NSW data to investigate whether selecting controls based on their predicted untreated outcomes can yield treatment effect estimates that closely approximate the experimental benchmark.\footnote{The availability of an experimental benchmark makes this comparison particularly informative, although the experimental estimate remains subject to sampling uncertainty and its causal interpretation relies on the absence of interference between units.}

We retain the DW sample of 185 treated men and use CPS-1, the largest CPS-based comparison sample considered by DW, comprising 15,992 nonexperimental controls, as our observational comparison group. We use the variables labeled \texttt{re74} and \texttt{re75} as pre-treatment earnings measures and \texttt{re78} as the evaluation outcome.\footnote{For NSW participants, \texttt{re74} measures earnings in months 13--24 before random assignment and does not necessarily correspond to calendar year 1974. For CPS controls, it refers to calendar-year earnings. The common variable label therefore does not ensure exact temporal alignment between samples. The historical forecasting model is estimated on CPS controls, for whom the 1974--1975 transition uses calendar-year earnings.} The 260 randomized NSW controls are used exclusively to construct the experimental benchmark and are excluded from both prognostic-score estimation and the construction of nonexperimental comparison groups.\footnote{The data files and variable definitions are distributed at \url{https://users.nber.org/~rdehejia/nswdata2.html}. All earnings amounts follow the units in those files.} The demographic covariates $Z_i$ comprise age, years of schooling, and binary indicators for Black race, Hispanic ethnicity, being married, and lacking a high school diploma.

For each set of control weights, we report two treatment effect estimates based on observed outcomes:
\begin{equation}
\widehat\tau_{\mathrm{Level}}=\Delta_{\omega}Y_{78},
\qquad
\widehat\tau_{\mathrm{DiD}}=\Delta_{\omega}(Y_{78}-Y_{75}),
\label{eq:nsw_observed_contrasts}
\end{equation}
where $\Delta_{\omega}$ is defined in Equation~\eqref{eq:treatment_effects}. Using the randomized NSW controls with equal weights, the estimated treatment effect is \$1,794.34 (SE = \$670.07) when comparing outcome levels and \$1,529.20 (SE = \$714.15) when applying DiD.\footnote{Their difference reflects a finite-sample pre-treatment earnings gap, not two distinct known treatment effects.} We compare each observational estimate with its corresponding experimental benchmark, ensuring that both are based on the same outcome measure and estimation approach.

\subsection{Prognostic score estimation}

We consider two alternative approaches to estimating prognostic scores. The first follows the conventional outcome-model implementation discussed by \citet{Hansen2008}: TabPFN~3.5 uses untreated CPS-1 controls as context to predict 1978 earnings based on demographic characteristics ($Z_i$), earnings in 1974 and 1975, and indicators for zero earnings in each of these years.\footnote{In the CPS-1 control sample, TabPFN~3.5 with four ensemble members achieves the highest out-of-sample $R^2$ and the lowest root mean squared error among the three methods considered: TabPFN, OLS, and random forest. Appendix~\ref{app:tabpfn_predictive_comparison} reports these predictive performance results.} To avoid using a control unit's outcome to predict itself, we implement three-fold cross-fitting. For each fold $k$, a model $\widehat g_{-k}$ is trained on the remaining two folds and used to generate out-of-fold predictions for the held-out controls, as well as predictions for all treated units. The two resulting prognostic scores used to construct the comparison groups are defined as follows:
\begin{equation}
s_{i,k}^{g,\mathrm{Level}}=\widehat g_{-k}(X_i),
\qquad
s_{i,k}^{g,\mathrm{Delta}}=\widehat g_{-k}(X_i)-Y_{i,75},
\label{eq:nsw_direct_scores}
\end{equation}
where $X_i$ includes the demographic characteristics $Z_i$ and the remaining predictors described above. Each control unit's prognostic score is based on its out-of-fold predicted 1978 earnings rather than its actual earnings in that year.

The second approach learns only from the historical transition between 1974 and 1975. On the same training-control folds, a separate TabPFN model estimates
\begin{equation}
Y_{i,75}=f\!\left(Z_i,Y_{i,74},\mathbf{1}\{Y_{i,74}=0\}\right)+u_i.
\label{eq:nsw_preonly_training}
\end{equation}
Applying this fitted relationship once to information at the later pre-treatment date gives
\begin{equation}
\begin{aligned}
\widehat p_{i,76,k}
&=\max\!\left\{0,\widehat f_{-k}\!\left(Z_i,Y_{i,75},\mathbf{1}\{Y_{i,75}=0\}\right)\right\},\\
s_{i,k}^{p,\mathrm{Level}}&=\widehat p_{i,76,k},
\qquad
s_{i,k}^{p,\mathrm{Delta}}=\widehat p_{i,76,k}-Y_{i,75}.
\end{aligned}
\label{eq:nsw_preonly_scores}
\end{equation}
No observed 1978 outcome enters these models or their score-based weights. Demographics are held fixed as pre-program descriptors because their annual histories are unavailable. Predictions are made for treated and held-out controls using the same fold-specific model.

The historical prognostic score in Equation~\eqref{eq:nsw_preonly_scores} predicts earnings in 1976, which are not observed in the available data. We use this one-step ahead forecast as a proxy for untreated earnings in 1978, rather than recursively forecasting subsequent years.\footnote{This introduces a horizon mismatch, as the model is trained on a one-year transition while the evaluation concerns outcomes three years after 1975. With a longer pre-treatment history, we could have trained the model to predict outcomes directly over a three-year horizon, aligning the forecasting task with the evaluation period. Given the available data, the approach requires the historical predictive relationship to remain sufficiently informative about untreated earnings in 1978. The resulting predictions are therefore used as prognostic scores, although their ability to capture expected untreated earnings in 1978 has not been established.}

\subsection{A common matching and reweighting design}

We estimate propensity scores using the logit specification from our replication of \citet{dehejia1999}, based on pre-treatment covariates of all 185 NSW treated units and 15,992 CPS-1 controls. The resulting propensity scores are used to construct comparison groups for both the levels and DiD estimators, with the same control weights applied to both outcomes within each weighting procedure.

We apply two procedures—one-to-one nearest-neighbor matching with replacement and entropy balancing (EB) of the first two score moments \citep{hainmueller2012}—to the propensity score and four prognostic scores: predicted untreated outcome levels and changes, each estimated either using post-treatment outcomes of untreated controls or exclusively pre-treatment data.\footnote{Entropy balancing offers some advantages over nearest-neighbor matching by directly enforcing balance on the selected score moments, whereas nearest-neighbor matching does not guarantee aggregate balance. Moreover, \citet{KingNielsen2019} show that conventional propensity-score matching can sometimes increase imbalance and model dependence.} To ensure comparability across methods, all procedures use the same three held-out control pools, including matching based on the propensity score estimated on the full sample. For nearest-neighbor matching, each treated unit is matched to its closest control within each pool $\mathcal C_k$, based on the corresponding score. For entropy balancing, uniform base weights $q_{j,k}$ are adjusted by solving
\begin{equation}
\begin{aligned}
\min_{\{\omega_{j,k}\geq0\}}\quad&
\sum_{j\in\mathcal C_k}\omega_{j,k}\log(\omega_{j,k}/q_{j,k}),\\
\text{subject to}\quad&\sum_{j\in\mathcal C_k}\omega_{j,k}=1,\\
&\sum_{j\in\mathcal C_k}\omega_{j,k}s_{j,k}^{\,r}
=\frac{1}{n_T}\sum_{i\in\mathcal T}s_{i,k}^{\,r},\qquad r=1,2.
\end{aligned}
\label{eq:nsw_entropy_balance}
\end{equation}
The control weights obtained separately within each held-out pool are normalized and then averaged, assigning equal weight to each pool. Under nearest-neighbor matching, every treated unit is matched to one control per pool, while all treated units retain equal weight in the estimation. For entropy balancing, scores are centered and scaled for numerical stability without altering the first- and second-moment balance constraints.

Importantly, all model specifications, fold assignments, and weighting procedures were fixed before comparing the estimates with the experimental benchmark. Appendix~\ref{app:implementation} provides further details on the replication and software settings.\footnote{As a replication check, using the full CPS-1 control pool reproduces the unadjusted nearest-neighbor estimate of \$1,582.46 reported by \citet{dehejia1999}. This differs from our main propensity-score matching estimate reported in Panel A of Table~\ref{tab:cps1_scores} because we use three separate control pools to ensure comparability across methods, while retaining the same propensity-score specification.}

\subsection{Treatment effect estimates and comparison with the experimental benchmark}

\begin{table}[!htbp]
\centering
\caption{CPS-1 comparisons with the experimental benchmark}
\label{tab:cps1_scores}
\small
\setlength{\tabcolsep}{4pt}
\begin{tabular}{@{}lrrr@{}}
\toprule
Comparison procedure & ATT: levels & ATT: DiD & Control ESS \\
\midrule
Randomized NSW controls & 1,794.34 (670.07) & 1,529.20 (714.15) & 260 \\
\addlinespace[2pt]
Unweighted CPS-1 controls & -8,497.52 (581.90) & 3,621.23 (609.85) & 15,992 \\
\midrule
\multicolumn{4}{@{}l}{\textit{A. Propensity score: replicated DW logit}} \\
Nearest neighbor & 1,990.77 (776.27) & 1,919.48 (740.88) & 79.70 \\
EB ($s$, $s^2$) & -452.48 (844.06) & 2,130.40 (823.12) & 99.78 \\
\addlinespace[5pt]
\multicolumn{4}{@{}l}{\textit{B. Outcome-trained prognostic scores: predicted 1978 level}} \\
Nearest neighbor & 1,651.17 (672.65) & 1,660.59 (683.84) & 323.90 \\
EB ($s$, $s^2$) & 1,422.98 (585.70) & 1,468.45 (614.25) & 3,770.88 \\
\addlinespace[5pt]
\multicolumn{4}{@{}l}{\textit{C. Outcome-trained prognostic scores: predicted earnings change (1975-1978)}} \\
Nearest neighbor & -3,546.99 (742.39) & 1,326.41 (712.06) & 368.01 \\
EB ($s$, $s^2$) & -3,677.55 (585.27) & 1,374.61 (612.45) & 7,275.86 \\
\addlinespace[5pt]
\multicolumn{4}{@{}l}{\textit{D. Pre-treatment-only prognostic scores: forecast 1976 level}} \\
Nearest neighbor & 1,135.40 (688.95) & 961.28 (697.96) & 269.96 \\
EB ($s$, $s^2$) & 684.51 (587.15) & 412.16 (616.09) & 3,716.91 \\
\addlinespace[5pt]
\multicolumn{4}{@{}l}{\textit{E. Pre-treatment-only prognostic scores: forecast earnings change (1975-1976)}} \\
Nearest neighbor & -1,668.32 (705.89) & 994.04 (706.80) & 349.63 \\
EB ($s$, $s^2$) & -2,234.32 (584.84) & 929.80 (612.88) & 6,601.11 \\
\bottomrule
\end{tabular}
\medskip
\begin{minipage}{\linewidth}
\footnotesize
\textit{Notes:} Dollar estimates use all 185 treated NSW units. Conditional sandwich standard errors, treating estimated scores and control weights as fixed, are reported in parentheses. The levels estimator uses observed 1978 earnings, while DiD uses the change in earnings between 1975 and 1978. The three folds used for cross-fitting prognostic scores also serve as common control pools across all comparison methods. Nearest-neighbor matching selects one control with replacement per treated unit per pool; normalized pool-specific designs are averaged equally. EB constrains the first and second moments of the scalar score. Panels B--C learn from 1978 outcomes of training controls; panels D--E use only pre-treatment training data and a one-step 1976 forecast as a proxy for 1978. Panel E subtracts 1975 earnings from that forecast. The propensity score is unchanged across outcome columns. Control ESS (effective sample size) measures the concentration of the final aggregate control weights. Appendix~\ref{app:implementation} gives exact specifications and inference details.
\end{minipage}
\end{table}

Table~\ref{tab:cps1_scores} reports ATT estimates based on two estimation approaches, outcome levels and DiD, using comparison groups constructed through nearest-neighbor matching or entropy balancing. Both procedures are applied to propensity scores and to prognostic scores estimated either from post-treatment outcomes of untreated controls or exclusively from pre-treatment information. When comparing outcome levels, the closest estimate to the experimental benchmark of \$1,794.34 is obtained through nearest-neighbor matching on the outcome-trained level prognostic score (\$1,651.17). Matching on the replicated DW propensity score also yields a nearby point estimate (\$1,990.77), while entropy balancing on the outcome-trained level score yields a somewhat lower estimate of \$1,422.98. Overall, these results suggest that prognostic-score matching can provide estimates closely aligned with the experimental benchmark, while traditional propensity-score matching also yields estimates close to the experimental reference.

Turning to DiD, the best agreement with the experimental benchmark of \$1,529.20 is obtained through entropy balancing on the outcome-trained level prognostic score, which yields an ATT of \$1,468.45. Interestingly, using the predicted-change score, which is more directly aligned with the DiD identification strategy, produces point estimates farther from the experimental reference: \$1,374.61 with entropy balancing and \$1,326.41 with nearest-neighbor matching. Conventional propensity-score matching yields a higher estimate of \$1,919.48. In this application, entropy balancing on predicted untreated levels yields the DiD point estimate closest to its experimental reference.

Results are less encouraging when prognostic scores are constructed exclusively from pre-treatment information. For the level comparison, nearest-neighbor matching on predicted 1976 earnings yields an ATT of \$1,135.40, while entropy balancing produces a lower estimate of \$684.51, both substantially below the experimental benchmark of \$1,794.34. For DiD, matching and entropy balancing on the predicted-change score yield estimates of \$994.04 and \$929.80, respectively, compared with the experimental benchmark of \$1,529.20. Interestingly, within the entropy-balancing approach, switching from predicted levels to predicted changes moves the DiD estimate closer to the experimental benchmark, from \$412.16 to \$929.80, although a substantial discrepancy remains. Overall, the stronger performance of outcome-trained prognostic scores does not extend to their exclusively pre-treatment counterparts in this application. However, these findings should be interpreted in light of the limited pre-treatment history and the reliance on predicted 1976 earnings as a proxy for untreated outcomes in 1978, rather than as evidence against the potential of historical forecasting more generally.

The negative treatment effect estimates based on outcome levels under predicted-change weights are not inconsistent with the positive DiD estimates. This difference reflects substantial imbalances in pre-treatment earnings between treated and weighted control units. For any given set of weights,

\begin{equation}
\widehat\tau_{\mathrm{Level}}-\widehat\tau_{\mathrm{DiD}}
=\Delta_{\omega}Y_{75}.
\label{eq:nsw_level_did_identity}
\end{equation}

As implied by the entropy-balancing estimates in Panels B and C of Table~\ref{tab:cps1_scores}, the treated-minus-weighted-control gap in 1975 earnings is approximately $-\$5{,}052$ when balancing predicted changes, compared with $-\$45$ when balancing predicted levels. Since the former approach targets similarity in expected untreated changes rather than outcome levels, its performance is more appropriately evaluated using DiD estimates. This illustrates how the choice of comparability criterion can substantially affect treatment effect estimates when pre-treatment outcome levels remain unbalanced.

\subsection{Weight concentration, balance, and the limits of the benchmark}

Agreement with the experimental benchmark should also be assessed in light of how control weights are distributed. To this end, we calculate the effective sample size of the control group, $\mathrm{ESS}_{C}=(\sum_j w_j)^2/\sum_j w_j^2$, using the final aggregate weights. Lower values indicate greater weight concentration, with fewer controls effectively contributing to the estimation. Outcome-trained level-score entropy balancing has ESS 3,770.88; balancing its change score raises ESS to 7,275.86. Propensity-score entropy balancing has ESS only 99.78 and produces less favorable point estimates under both outcomes. The propensity nearest-neighbor design also has relatively concentrated weights (ESS 79.70), but its level estimate remains close to the experimental reference.

Across the ten pre-treatment predictors, including the two derived zero-earnings indicators, propensity-score matching achieves a mean absolute standardized difference (MASMD) of 0.026, compared with 0.522 for nearest-neighbor matching on the outcome-trained level prognostic score. Nevertheless, the latter produces a level point estimate closer to the experimental reference. This comparison illustrates that better measured covariate balance need not coincide with greater proximity to an experimental point estimate. Such proximity can reflect sampling variation as well as differences in bias.

Overall, outcome-trained prognostic scores yield the estimates closest to the experimental benchmarks, with nearest-neighbor matching closest to the level benchmark and entropy balancing closest to the DiD benchmark. However, this empirical application is intended primarily as a proof of concept, illustrating the feasibility of using predicted untreated outcomes to construct comparison groups rather than establishing the superiority of any particular method. The findings should also be interpreted in light of the sampling uncertainty surrounding the experimental benchmarks. The exclusively pre-treatment approach yields less encouraging results, although its evaluation is constrained by the limited historical information available and the use of a one-year forecast as a proxy for outcomes three years ahead. Further evidence from applications with longer pre-treatment histories and appropriately aligned forecasting horizons is therefore needed to assess the broader potential of the proposed framework.

\section{Conclusions}

This paper proposes a forward-looking approach to constructing comparison groups in observational causal inference. Rather than relying exclusively on similarities in observed pre-treatment characteristics or historical outcome trajectories, we identify suitable controls based on their predicted outcomes in the absence of treatment during the post-treatment period. The relevant prediction target depends on the causal design: expected outcome levels for comparisons in levels and expected outcome changes for DiD. Crucially, predictions guide the selection or weighting of controls, while their observed outcomes remain the basis for estimating treatment effects. This distinction allows forecasting errors shared by treated and control units to cancel out, although differences in their prediction errors remain a threat to causal identification.

Revisiting the NSW benchmark illustrates both the potential and the limitations of this approach. Prognostic scores trained on observed outcomes of untreated controls yield treatment effect estimates close to the experimental benchmarks, with nearest-neighbor matching performing best for outcome levels and entropy balancing for DiD. Importantly, better pre-treatment covariate balance, more accurate alignment of the score with the estimation strategy, and larger effective control samples do not necessarily translate into estimates closer to the experimental benchmarks. These findings reinforce a central message: similarity in observed pre-treatment characteristics does not guarantee similarity in untreated potential outcomes, and the quality of a comparison group cannot be established through any single balancing diagnostic.

The results are less encouraging when prognostic scores are constructed exclusively from pre-treatment data. However, this implementation relies on a one-year forecast as a proxy for outcomes three years ahead, limiting what can be concluded about the broader potential of historical forecasting. Applications with longer pre-treatment histories could allow more reliable predictions over the relevant evaluation horizon, together with temporal validation and placebo exercises. More generally, advances in TFMs offer new opportunities to exploit available information when constructing counterfactual comparisons. Their contribution to causal inference, however, lies not in replacing identifying assumptions, but in helping researchers construct comparison groups that more closely reflect the untreated outcomes they seek to recover.

\clearpage
\appendix
\section{Implementation details}
\label{app:implementation}

\subsection{Prediction and score construction}

All prognostic models use TabPFN package 9.2.0, model version \texttt{V3\_5}, checkpoint \texttt{tabpfn-v3.5-20260909.safetensors} \citep{priorlabs2026}, and four ensemble members. The maximum training context is 2,048 observations per member; the full training fold supplies the candidate context rows. The 15,992 CPS-1 controls are assigned to three approximately equal folds using seed 12345. Model $k$ uses seed $12345+k$, is fitted on the other two control folds, and predicts all 185 treated units and the held-out controls. The same fold assignments and candidate donor pools are retained for every score and weighting method. Negative earnings predictions are set to zero.

The outcome-trained model predicts 1978 earnings using ten pre-treatment predictors: age, schooling, Black and Hispanic indicators, marital status, absence of a high-school diploma, earnings in 1974 and 1975, and indicators for zero earnings in each year. The model is trained on observed 1978 earnings of untreated CPS-1 controls.

The model trained exclusively on pre-treatment data instead uses eight predictors: the same six demographic characteristics, lagged earnings, and an indicator for zero lagged earnings. Specifically, it learns to predict 1975 earnings using demographic characteristics and earnings observed in 1974. The estimated relationship is then applied to information available in 1975 to forecast earnings in 1976. This forecast serves as a proxy for expected untreated earnings in 1978, without using any observed 1978 outcomes to train the model or construct comparison weights. Demographic characteristics are held fixed across the two periods because annual measurements are unavailable. Finally, the predicted-change score is obtained by subtracting observed 1975 earnings from the forecast for 1976.

To ensure reproducibility, all saved predictions are checked against the input variables, training outcomes, fold assignments, random seeds, and model settings. These checks also ensure that introducing models trained exclusively on pre-treatment data does not alter the results obtained with outcome-trained prognostic scores. Neither the outcomes of randomized controls nor the 1978 earnings of treated units are used to estimate prognostic scores.

\subsection{Propensity specification and common donor pools}

The propensity-score model follows the logit specification used in our replication of \citet{dehejia1999}. It includes age and its square and cube, schooling and its square, marital status, Black and Hispanic indicators, earnings in 1974 and 1975, indicators for zero earnings in each year, and an interaction between schooling and 1974 earnings. The no-degree indicator is excluded, and earnings variables are rescaled for numerical stability. The model is estimated with a convergence tolerance of $10^{-12}$ and a maximum of 100 iterations. Using the full CPS-1 control pool for nearest-neighbor matching reproduces the unadjusted ATT estimate of \$1,582.46 reported in Table~3, column~7 of \citet{dehejia1999}, to the published precision. This replication concerns the point estimate and does not imply that the original matching assignments or standard errors have been recovered.

Our main analysis uses the same three control pools across all methods to ensure comparability. Within each pool, nearest-neighbor matching selects the control with the closest score for each treated unit, allowing controls to be matched multiple times. Ties are resolved according to the original order of observations in the dataset. Entropy balancing instead assigns weights to control units to match the first two moments of the score distribution among treated units, using the \texttt{WeightIt} implementation with uniform initial weights. Balance constraints are verified numerically after normalization. No additional restrictions are imposed on individual covariates. Consequently, entropy balancing on the propensity score should not be confused with conventional inverse-odds weighting for the ATT.

Control weights are normalized to sum to 185 within each pool and then averaged across pools, giving each pool equal importance. All 185 treated units retain a weight of one, while the final control weights also sum to 185. To compute the treatment effect estimates in Equation~\eqref{eq:nsw_observed_contrasts} in the main text, control weights are divided by 185. This normalization does not affect the effective sample size (ESS), which accounts for the concentration of weights. In particular, controls selected multiple times accumulate weight, so the ESS may be substantially smaller than the number of distinct controls contributing to the estimation.

\subsection{Treatment effect estimation, inference, and verification}

All reported ATT estimates are obtained as differences in observed weighted means, using either 1978 earnings or changes in earnings between 1975 and 1978. No additional outcome regression or covariate adjustment is applied. Conditional sandwich standard errors are computed using the estimated weights and assuming independence across units. These standard errors treat the estimated scores and resulting comparison weights as fixed, thus excluding uncertainty arising from score estimation, control selection, and entropy balancing. Moreover, because observational estimates and experimental benchmarks share the same treated units, their reported standard errors cannot be treated as independent when assessing differences between estimates.

We independently verify all ATT estimates as differences in weighted means and reproduce their conditional standard errors using the corresponding influence functions. Additional checks confirm the separation between training and held-out observations, the accuracy of nearest-neighbor assignments, the normalization of control weights, and the satisfaction of entropy-balancing constraints. We also verify the identity in Equation~\eqref{eq:nsw_level_did_identity} in the main text and confirm that the models trained exclusively on pre-treatment data use only eligible historical information. Finally, introducing these additional models leaves the original outcome-trained prognostic-score and propensity-score estimates numerically unchanged.

\section{Out-of-sample predictive performance in the CPS-1 control sample}
\label{app:tabpfn_predictive_comparison}

We compare TabPFN~3.5, ordinary least squares (OLS), and a random forest in two prediction tasks using the 15,992 untreated CPS-1 observations. The historical task predicts 1975 earnings using age, schooling, indicators for Black race, Hispanic ethnicity, being married, and lacking a high school diploma, together with 1974 earnings and an indicator for zero earnings in 1974. The second task predicts 1978 earnings using the same demographic covariates, earnings in both 1974 and 1975, and indicators for zero earnings in each pre-treatment year. Thus, the two tasks contain eight and ten predictors, respectively.

All methods are evaluated using the same three fixed control-sample folds and the same predictors within each task. Models are fitted using the other two folds, and each observation receives a prediction only when held out. No held-out outcome is used to fit the model producing that observation's prediction. We report pooled out-of-sample performance across all three evaluation folds, rather than an average of fold-specific $R^2$ values. Specifically,
\begin{equation}
R^2_{\mathrm{OOS}}
=1-\frac{\sum_{i\in\mathcal C}(Y_{i,t}-\widehat Y_{i,t}^{\mathrm{OOF}})^2}
{\sum_{i\in\mathcal C}(Y_{i,t}-\overline Y_{\mathcal C,t})^2},
\end{equation}
where $\mathcal C$ denotes the CPS-1 control sample, $\widehat Y_{i,t}^{\mathrm{OOF}}$ is the held-out prediction, and $\overline Y_{\mathcal C,t}$ is the observed sample mean for the target year. Root mean squared error (RMSE) is the square root of the mean squared held-out prediction error and is expressed in dollars.

Table~\ref{tab:tabpfn_predictive_comparison} reports the out-of-sample predictive performance of the alternative models. TabPFN consistently achieves the highest $R^2$ and lowest RMSE across both prediction tasks, outperforming OLS and random forests. Relative to OLS, the RMSE reductions are 0.34\% for 1975 and 0.83\% for 1978. These improvements are achieved in a setting characterized by a limited number of predictors and a short pre-treatment history, suggesting that richer datasets may offer greater opportunities to exploit TabPFN's predictive capabilities. The reported evaluations assess predictive accuracy among held-out CPS controls within each target year. Extending this performance to forecasts for subsequent years and to treated units requires temporal stability and comparability of prediction errors across groups, which remain central considerations for the proposed causal framework.

\begin{table}[!htbp]
\centering
\caption{Out-of-sample earnings prediction for CPS-1 controls}
\label{tab:tabpfn_predictive_comparison}
\small
\begin{tabular}{llrr}
\toprule
Target year & Prediction method & RMSE & OOS $R^2$ \\
\midrule
1975 & TabPFN~3.5 & \textbf{4,522.38} & \textbf{0.7620} \\
1975 & OLS & 4,537.96 & 0.7604 \\
1975 & Random forest & 4,686.17 & 0.7445 \\
\midrule
1978 & TabPFN~3.5 & \textbf{6,925.47} & \textbf{0.4846} \\
1978 & OLS & 6,983.19 & 0.4760 \\
1978 & Random forest & 7,044.56 & 0.4668 \\
\bottomrule
\end{tabular}
\medskip
\begin{minipage}{\linewidth}
\footnotesize
\textit{Notes:} All results use out-of-fold predictions for the same 15,992 CPS-1 controls. The best value within each prediction task is in bold. TabPFN uses four ensemble members with at most 2,048 training observations per member; OLS and random forest use the full training folds. OLS includes additive terms for the listed predictors. Random forests use 500 trees, bootstrap sampling, minimum node size five, and two candidate predictors per split for the 1975 task or three for the 1978 task. Settings are fixed without tuning on the evaluation folds. Negative predictions are set to zero for every method. RMSE is in dollars. OOS $R^2$ is computed from pooled held-out prediction errors using the formula above.
\end{minipage}
\end{table}

\end{document}